# Unified definition of ferroelectricity

Wei Luo[1]*, Sihan Deng[2]*, Hongjun Xiang[2]† and Laurent Bellaiche[1,3]

[1]Smart Ferroic Materials Center, Physics Department and Institute for Nanoscience and Engineering, University of Arkansas Fayetteville, Arkansas 72701, USA

[2]Key Laboratory of Computational Physical Sciences (Ministry of Education), State Key Laboratory of Surface Physics, and Department of Physics, Fudan University, Shanghai 200433, China

[3]Department of Materials Science and Engineering, Tel Aviv University, Ramat Aviv, Tel Aviv 6997801, Israel

*These authors contributed equally to this work.

†Contact author: hxiang@fudan.edu.cn

**Abstract:**

Recent theoretical and experimental advances in quantum ferroelectrics suggest that ferroelectricity can also emerge in non-polar space group, highlighting the limitations of conventional polar space group criteria in identifying ferroelectric materials. Here, we introduce a unified definition based on switchable polarization differences between energetically equivalent states, which naturally encompasses conventional and quantum ferroelectrics. Guided by this principle, we implement a high-throughput screening strategy that systematically identifies both conventional and quantum ferroelectrics among experimentally synthesized materials. In particular, we identify a new type of quantum ferroelectric in which the quantized polarization arises from arbitrary ionic displacements, in contrast to previous quantum ferroelectrics (including both fractional and integer quantum ferroelectrics) where quantized polarization results from fractional or integer ionic displacements. Notably, we find that materials such as $Ba_3I_6$ and $Cs_2PdC_2$ exhibit low switching barriers and robust insulating behavior, highlighting their experimental viability. Our results reconcile conventional and quantum ferroelectrics, expand the accessible materials landscape, and provide a practical roadmap for discovering next-generation ferroelectrics with advanced switchable functionalities.

## Introduction

The classical definition of ferroelectricity states that a material is ferroelectric if it possesses two or more equivalent states with *spontaneous polarization*, and these states can be reversibly switched by an external electric field[1]. Such switchable polarization is essential for memory and logic applications, while the associated non-centrosymmetric lattice and strong polarization–strain coupling further enable piezoelectric sensing, actuation, and energy conversion technologies[2-11]. Foundational advances in the modern theory of polarization have enabled predictive, first-principles design and screening of ferroelectric and multiferroic materials across diverse chemistries and structures[12-14]. Recent high-throughput workflows have further accelerated the discovery of device-relevant candidates with low coercive fields and robust endurance[15-17]. Together, these developments have propelled the field beyond classical perovskite oxides to encompass low-dimensional, van der Waals and intercalated systems, opening routes to scalable ferroelectric devices[18-23].

Concurrently, theory and experiment have revealed quantum ferroelectricity (QFE) that transcend the classical picture of small, symmetry-breaking displacements confined to polar space groups. Prototypical 2D van der Waals ferroelectrics such as $\alpha$-$In_2Se_3$ exhibit interlocked in-plane/out-of-plane switching ("dipole locking"), robust ferroelectricity down to a few-nanometer thickness, and room-temperature operation[18,24-27]. Because its $C_{3v}$ symmetry nominally permits only out-of-plane polarization, this behavior highlights the unconventional ferroelectric character of $\alpha$-$In_2Se_3$. Intercalated and layered chalcogenides (e.g., $CuCrS_2$/$CuCrSe_2$/$AgCrS_2$) feature large-amplitude cation displacements and intercorrelated polar axes, with both lateral and vertical switching pathways captured in devices[28-31]. Integer and fractional quantum ferroelectricity (FQFE)[32] in crystals of nominally nonpolar or high symmetry, demonstrate that measurable polarization changes can be integer or fractional multiples of the polarization quantum[33-36], challenging conventional, purely symmetry-based definitions. Beyond these systems, long-displacement–induced ferroelectrics can also exhibit quantized polarization[37-47]. These discoveries have created a conceptual schism, where a growing class of materials exhibits the essential functional characteristic of

ferroelectricity—electrically switchable states—yet falls outside its formal definition. This situation calls for a fundamental re-evaluation and a new, unified framework that can consistently describe both conventional and these emergent quantum ferroelectrics (QFEs).

Here, we address this fundamental challenge by proposing a unified definition for ferroelectricity: *Ferroelectricity refers to a class of materials that possess two or more energetically equivalent states with a nonzero polarization difference between them, and these states can be reversibly switched by an external electric field*. This definition generalizes *spontaneous polarization* by emphasizing the physically measurable quantity in ferroelectric switching, namely the *polarization difference* between two switchable states. In the modern theory of polarization, the absolute value of polarization under periodic boundary conditions is multivalued and depends on the choice of reference branch[12,48], whereas the polarization difference between two insulating states is well defined and experimentally meaningful[12,49,50]. From this perspective, spontaneous polarization relative to an appropriate reference state remains a useful and often intuitive concept for conventional ferroelectrics, but it may be insufficient or ambiguous for QFEs. Therefore, our definition should be understood not as a rejection of the traditional framework, but as its natural extension: conventional ferroelectrics represent a special case in which the polarization difference can be conveniently described in terms of a spontaneous polarization vector. This generalized approach provides a consistent framework for unifying conventional and QFE. Based on this new principle, we developed a high-throughput screening strategy to identify both conventional and QFEs, enabling accelerated materials discovery. For conventional ferroelectrics, we identify 73 previously unreported candidates that were not captured in earlier high-throughput studies[15,17]. For QFEs, we discover a distinct class in which the quantized polarization arises from arbitrary ionic displacements, in contrast to previously reported QFEs where quantization is associated with fractional or integer lattice translations, thereby demonstrating the greater generality of our new definition of ferroelectricity. Notably, several of the candidates crystallize in non-polar space groups and exhibit low switching barriers, making them promising QFE candidates for experimental realization. Our work not only provides a unified theoretical framework that reconciles all known types of ferroelectricity but also delivers a concrete roadmap for the experimental discovery of novel ferroelectricity with advanced functionalities.

**Polarization differences as the fundamental criterion.**

The requirement of spontaneous polarization imposes strict symmetry constraints: only materials belonging to polar space groups can be ferroelectric, in accordance with the Neumann principle[51]. Symmetry constraints are highly effective for conventional ferroelectrics, in which spontaneous polarization arises from small ionic displacements (Top panel in Fig. 1a). However, recent experimental and theoretical advances have uncovered a distinct class of QFEs, including FQFE and integer quantum ferroelectricity (IQFE)[35,37,41,52]. In these systems, polarization originates from ionic displacements that are fractional or integer multiples of the lattice constant (Middle and bottom panels in Fig. 1a). These findings call into question whether spontaneous polarization is a necessary condition in the definition of ferroelectricity.

From the perspective of the modern theory of polarization, polarization in crystalline solids is a multivalued quantity, defined modulo a polarization quantum. Consequently, for an appropriate choice of reference structure, a nonzero polarization can, in principle, be assigned to a crystal of any symmetry. Moreover, under periodic boundary conditions, the absolute value of polarization is not uniquely defined, whereas polarization differences between insulating states are well defined, physically meaningful, and experimentally accessible. These considerations suggest that spontaneous polarization, although highly useful in the description of conventional ferroelectrics, is not always the most general or unambiguous criterion for identifying ferroelectric behavior. Motivated by this insight, we propose a unified definition of ferroelectricity that consistently encompasses both conventional and QFE. (Fig. 1b): ferroelectricity refers to a class of materials that possess two or more energetically equivalent states with a nonzero polarization difference between them, and these states can be reversibly switched by an external electric field. Unlike the traditional definition, this formulation places no symmetry restrictions on the material; instead, it emphasizes the polarization difference between equivalent states as the fundamental criterion. This shift is natural within the modern theory of polarization, where polarization differences between states are the central physical quantities. Based on this unified definition, we demonstrate how a high-throughput framework can be developed to systematically identify a broad class of ferroelectric materials, including both conventional and QFE.

**High-throughput screening for conventional ferroelectrics and QFEs.**

Based on the unified definition of ferroelectricity—namely, that a material is ferroelectric if it possesses two or more equivalent states with a finite polarization difference that can be switched by an external electric field—two essential conditions must be satisfied. First, the equivalent states must exhibit a nonzero polarization difference. Second, these states must be switchable under an external electric field, which implies that the energy barrier cannot be excessively large.

To satisfy the first condition, we start from an initial structure, denoted as $L1$ (Fig. 2a), with space group $G_L$ and corresponding symmetrized lattice symmetry $G_{Latt}$ ($G_{Latt}$ represents the space group of the lattice after removing all atoms). We then construct a final structure $L2$ such that $L2{=}hL1$, where $h$ is a symmetry operation belonging to $G_{latt}/G_L$ but not to $G_L$. The operation $h$ may correspond to a rotation (proper or improper), a pure translation, or a combined rotation–translation operation. If $h$ is a rotational operation (or a combined rotation–translation operation)—taking $M_{xy}$ as an example (see Fig. 2a,b)—The polarization difference between the L1 and L2 is generally unconstrained and can take arbitrary values, including fractional or integer multiples of the polarization quantum. In the general case (Fig. 2a,b), the ferroelectric states L1 and L2 are low-symmetry phases, where ions (red open circles) occupy low-symmetry positions. As a result, the polarization difference between L1 and L2 can vary continuously, as in conventional ferroelectrics. In contrast, when the states are high-symmetry phases, the ions occupy high-symmetry positions (red filled circles in Fig. 2a,b, with fractional displacement (1/2, 1/2)), which constrains the polarization difference to be only fractional or integer multiples of the polarization quantum. This is more clearly illustrated by the polarization lattice in Fig. 2c, which shows fractional multiples of the polarization quantum as an example. The purple filled circles and green dashed circles represent the polarizations of L1 and L2, respectively, and their difference corresponds to the vectors connecting points of different colors. Therefore, the polarization difference can be either fractionally or integer-quantized, depending on whether the ionic charges ($n$) are odd or even. In contrast, if $h$ is a pure translational operation (taking $T_{\frac{1}{2}y}$ as an example; see Fig. 2d,e), the polarization difference can only be an integer multiple of the polarization quantum, as is readily seen in Fig. 2f, where the purple filled circles and dashed green circle represent the polarization lattices of L1 and L2, respectively. After identifying the initial and final states, we employ Crystmatch[53,54] to determine the possible transition paths connecting them. The polarization difference along each path is then estimated using a point-charge model, and only cases with a finite polarization difference are retained. To satisfy the second condition, we perform nudged elastic band (NEB) calculations utilizing the solid-state NEB method[55] as implemented in TSASE, to determine the energy barriers between the initial and final states, retaining only those transitions with barriers below 400 meV per formula unit.

Our high-throughput screening workflow is summarized in Fig. 2g. Based on the Materials Project database[56], we selected materials with experimentally reported crystal structures, band gaps exceeding 0.1 eV, fewer than 50 atoms per unit cell, and no more than four distinct atomic species. Symmetry operations involving point-group and translational transformations are then applied to generate the corresponding final structures (L2). For transition paths exhibiting a finite polarization difference, NEB calculations are carried out, with total energies predicted by a universal machine-learning potential (UMLP), specifically the UMA model[57]. Using our high-throughput framework, we identify 100 conventional ferroelectrics and 68 QFEs, with energy barriers below 400 meV/f.u.. Specifically, for QFEs in the rotational case—where the initial and final states are related by a rotational symmetry operation—we identify: (i) 48 candidates belonging to nonpolar space groups, among which only 17 have been reported by previous methods[33] (see Table S1 of the Supplementary Material (SM)); and (ii) 12 candidates that simultaneously exhibit conventional ferroelectricity and QFE behavior, with only 4 of them previously identified (see Table S2 of the SM). For the translational case, we identify 8 candidates whose polarization differences are integer multiples of the polarization quantum (see Table S3 of the SM). Notably, none of these cases were captured by previous methods, as translational symmetry-related QFE was not considered in earlier approaches[33]. Remarkably, through our high-throughput approach based on the unified definition of FE, we identify a new type of QFE in both rotational and translational cases, in which the ionic displacements can take arbitrary values, as illustrated in Fig. 1c. Representative material candidates include $Cs_2PdC_2$ (rotational case, see discussion below) and $Ag_2I_2$ (translational case, part I of the SM). For the 100 conventional ferroelectrics, we find that only 27 were previously identified by theory and experiments [15,17], as highlighted in yellow in Table S4 of the SM. In Fig. 2h, we summarize the barriers, polarization difference and band gaps of all QFE candidates with barriers below 400 meV/f.u.. Note that the energy barriers shown in Fig. 2h were calculated using the NEB method based on the UMA model. A benchmark comparison between the density functional theory (DFT) and UMA results is provided in Fig. S1 of the SM, demonstrating good agreement between the two approaches. All classes of conventional ferroelectrics and QFEs and their associated properties—including space groups, band gap, polarization and barriers—are summarized in Table S1-S4 of the SM. In the following, we focus on representative candidates with low energy barriers to elucidate their ferroelectric properties and assess their experimental feasibility.

**$Ba_3I_6$: An IQFE in a non-centrosymmetric, nonpolar space group.**

We identify hexagonal alkaline-earth halides as an unexpected platform for switchable IQFE. As a representative example, we consider $Ba_3I_6$, which crystallizes in the non-centrosymmetric yet nonpolar space group P-62m ($G_L$). Upon removing all atoms, the corresponding lattice space group $G_{Latt}$ is P6/mmm. The top and side views of two symmetry-equivalent structural states, denoted L1 and L2, which are related by inversion symmetry, are shown in Fig. 3a (Left and right panels). Berry-phase calculations[12] reveal a polarization difference between L1 and L2 of approximately 21.6 $\mu C/cm^2$ (Fig. S2a of SM), corresponding to one polarization quantum. This polarization difference is oriented along the out-of-plane $z$ direction (the crystallographic $c$ axis). A comparison of the two states shows that the polarization difference originates from a translation of Ba-1 atom (highlighted by the blue sphere in Fig. 3a) by half a lattice vector along the $+z$ direction. Owing to the nominal +2 valence of Ba, this displacement gives rise to a polarization change equal to one polarization quantum. Although the polarization difference is solely determined by the displacement of the Ba-1 atom, this does not imply that the remaining atoms are static. In particular, the six iodine (I) atoms undergo small but correlated positional relaxations. In the L1 structure, the Ba-1 atom forms three short in-plane bonds (3.49 Å) with I-1, I-2, and I-3 (Left panel of Fig. 3a), and three longer out-of-plane bonds (3.84 Å) with I-4, I-5, and I-6. Upon the ferroelectric transition to the L2 structure, the Ba-1 atom is translated by half a lattice vector along the $c$ axis. Concomitantly, the iodine sublattice rearranges: I-4, I-5, and I-6 move away from one another within the basal plane, while I-1, I-2, and I-3 move closer together (Right panel of Fig. 3a). As a result, Ba-1 forms three short in-plane bonds (3.49 Å) with I-4, I-5, and I-6 and three longer out-of-plane bonds (3.84 Å) with I-1, I-2, and I-3, in contrast to the bonding configuration in the L1 state.

Using NEB calculations at the DFT level, we determine the energy barrier separating the L1 and L2 states, as shown in the middle panel of Fig. 3a. Notably, the energy profile exhibits no polar instability, in contrast to conventional perovskite ferroelectrics[58,59], and instead resembles the behavior reported for the QFE $CuCrSe_2$[31]. We find a low polarization-switching barrier of approximately 26 meV per formula unit. The coercive field as a function of temperature, calculated from molecular dynamics (MD) simulations based on a machine learning potential (MLP) (see method part for details), is shown in Fig. 4a. It can be seen that the coercive field is approximately 0.04 V/Å, slightly lower than the 0.05 V/Å reported for $PbTiO_3$ from MLP[60], suggesting that it should be experimentally accessible. Note that theoretical calculations often yield coercive fields about an order of magnitude higher than experiments[61], likely due to the Landauer paradox[62]. This difference may stem from experimental inhomogeneities, while calculations assume a defect-free medium. The band gap calculated using the Heyd-Scuseria-Ernzerhof (HSE) hybrid functional[63] is approximately 4.0 eV, confirming its robust insulating character, which is highly favorable for sustaining ferroelectric functionality. A closely related compound, $Ba_3Cl_6$, displays an analogous switching mechanism but with a substantially higher barrier of 101 meV per formula unit, underscoring the tunability of the switching energetics via chemical substitution. Together, these results identify alkaline-earth halide frameworks as a previously overlooked class of low-barrier QFEs and extend the design space for ferroelectric-like functionality beyond oxide-based systems. Other ferroelectric materials crystallizing in nonpolar space groups are summarized in Table S1 of the SM.

**$Cs_2PdC_2$: An IQFE in a centrosymmetric, nonpolar space group.**

In contrast to $Ba_3I_6$, $Cs_2PdC_2$ adopts a centrosymmetric crystal structure belonging to space group P-3m1 (No. 164; point group $D_{3d}$). As illustrated in Fig. 3b, the two configurations L1 and L2 are related by mirror symmetry ($M_z$) and are therefore energetically and structurally equivalent. Berry-phase calculations [12] yield a polarization difference of approximately 56.2 $\mu C/cm^2$ (Fig. S2b of SM) between these two states, oriented along the out-of-plane $z$ direction, corresponding to a single polarization quantum. A detailed comparison of the two configurations indicates that the polarization difference arises from distinct displacements of the two inequivalent cesium ions, denoted Cs-1 and Cs-2 in the left panel of Fig. 3b. Specifically, the L2 configuration can be generated from L1 by translating Cs-1 and Cs-2 by 0.45 (d1 in Fig. 3b) and 0.55 (d2 in Fig. 3b) lattice units along the positive $z$ (crystallographic $c$) direction, respectively, while the $PdC_2$ framework remains intact. Because both Cs-1 and Cs-2 carry a nominal +1 valence, the resulting polarization difference between L1 and L2 in $Cs_2PdC_2$ is exactly equal to one polarization quantum. Note that this is a newly identified nontrivial example of an IQFE in which the atomic displacements can take arbitrary values, while the polarization difference is exactly equal to one polarization quantum; to our knowledge, such a case has not been reported previously.

The energy barrier between L1 and L2 is estimated to be 78 meV per formula unit, as shown in middle panel of Fig. 3b. The coercive field as a function of temperature is shown in Fig. 4b, which is slightly lower than that of $Ba_3I_6$. HSE functional calculations indicate a band gap of approximately 2.6 eV, confirming its insulating behavior. Analogous compounds, $Cs_2PtC_2$ and $Rb_2PdC_2$, display similar ferroelectric properties, with energy barriers of roughly 82 meV and 131 meV, respectively (see Table S1 in the SM). Owing to their modest barriers and good insulating characteristics, we anticipate that experimental studies will soon be able to verify these intriguing ferroelectric behaviors.

In addition to $Ba_3I_6$ and $Cs_2PdC_2$, whose L1 and L2 states are related by a rotation operation, we also present two other candidates. One is $Ag_2I_2$, an IQFE belonging to a polar space group, in which the L1 and L2 states are related by a translation operation. The other is $Na_4Pd_2S_4$, an overlooked conventional ferroelectric in a polar space group. Details of these two candidates are provided in part I (Fig. S3) and part II (Fig. S4) of the SM.

## Conclusion

In conclusion, we propose a unified definition of ferroelectricity based on switchable polarization differences between energetically equivalent insulating states, rather than on spontaneous polarization constrained by polar space-group symmetry. This framework naturally reconciles conventional ferroelectrics with emerging QFEs and provides a general criterion for identifying switchable functionality in polar and nonpolar crystals. Guided by this principle, we develop a high-throughput screening workflow and identify 100 conventional ferroelectrics and 68 QFEs among experimentally synthesized materials, including many candidates missed by previous approaches. Notably, we uncover a distinct type of QFE in which quantized polarization arises from arbitrary ionic displacements, beyond previously known fractional or integer displacement mechanisms. Representative compounds such as $Ba_3I_6$ and $Cs_2PdC_2$ exhibit low switching barriers, robust insulating gaps, and experimentally accessible coercive fields, highlighting their promise for realization. Our results broaden the materials landscape of ferroelectricity and provide a practical roadmap for discovering next-generation switchable quantum materials.

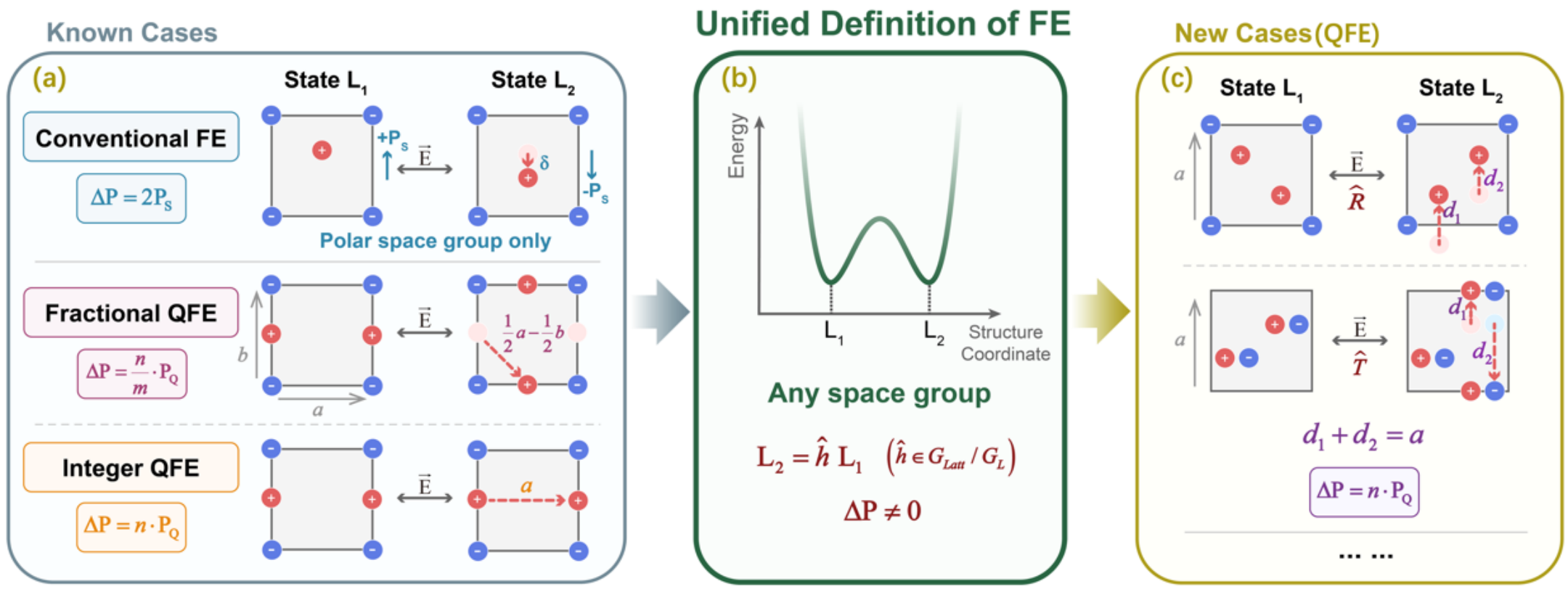

Figure 1 | (a) The Schematic diagram for conventional and QFE (FQFE and IQFE) ferroelectric mechanisms. (b) Unified ferroelectricity framework. (c) New type of QFE revealed by high-throughput screening.

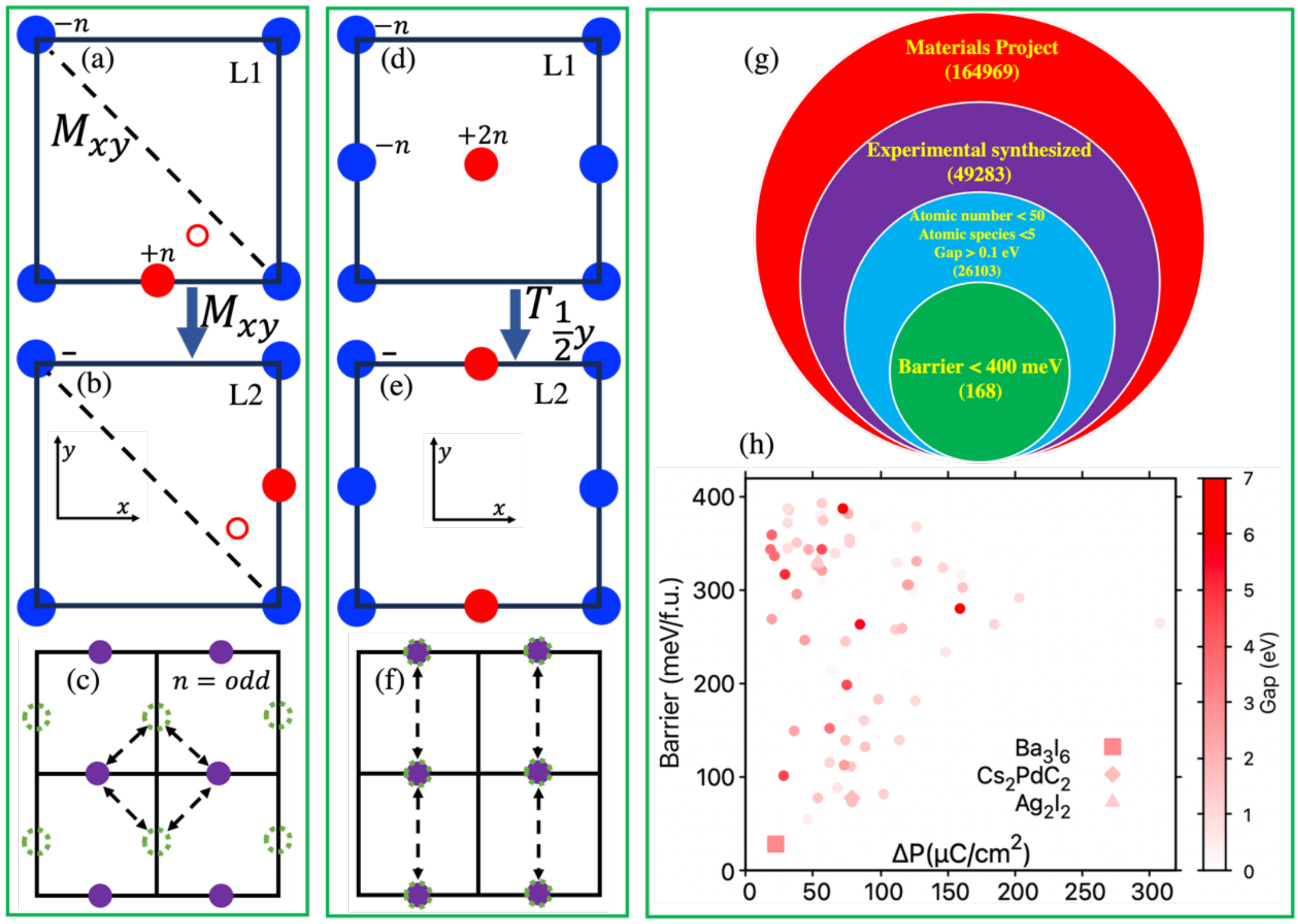

Figure 2 | The initial (a) and final (b) states for the rotation case. (c) Polarization lattices corresponding to (a) (purple) and (b) (green) for odd values of $n$. The initial (d) and final (e) states for the translation case. (f) Polarization lattices corresponding to (d) (purple) and (e) (green). (g) High-throughput screening workflow. (h) Energy barriers, polarization difference, and band gap for QFE candidates. The values of band gaps are extracted from the Materials Project [56].

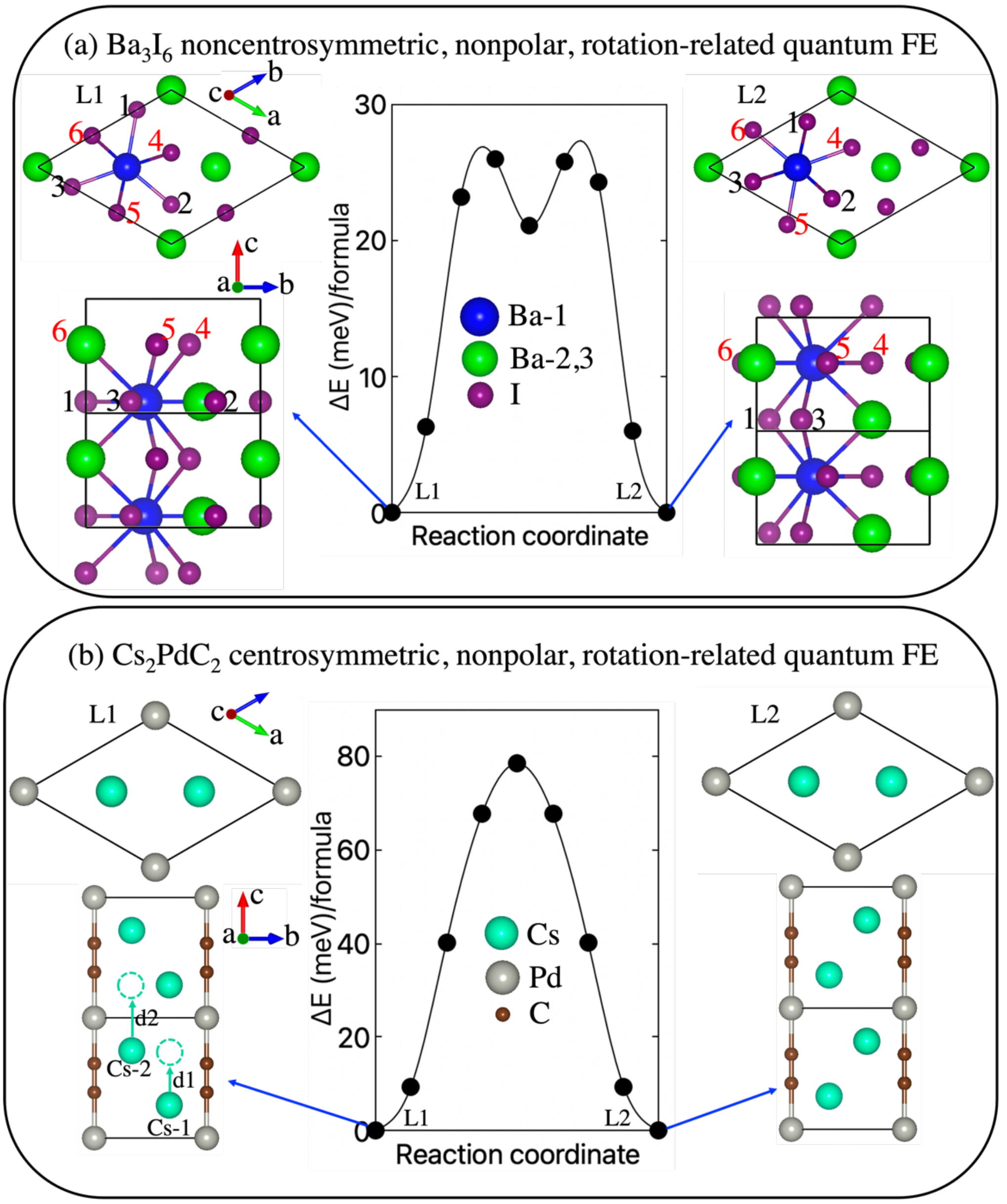

Figure 3 | (a) Crystal structure and migration pathway of $Ba^{2+}$ ions in $Ba_3I_6$. The calculated energy profile along the migration pathway from L1 to L2 reveals the diffusion barrier of $Ba^{2+}$ ions, featuring two symmetric energy maxima of approximately 28 meV/f.u.. (b) Crystal structure and migration pathway of $Cs^{+}$ ions in $Cs_2PdC_2$. The distances d1 and d2 in left panel indicate the migration pathway of Cs-1 and Cs-2. The energy profile along the L1–L2 migration pathway exhibits a barrier of ~78 meV/f.u..

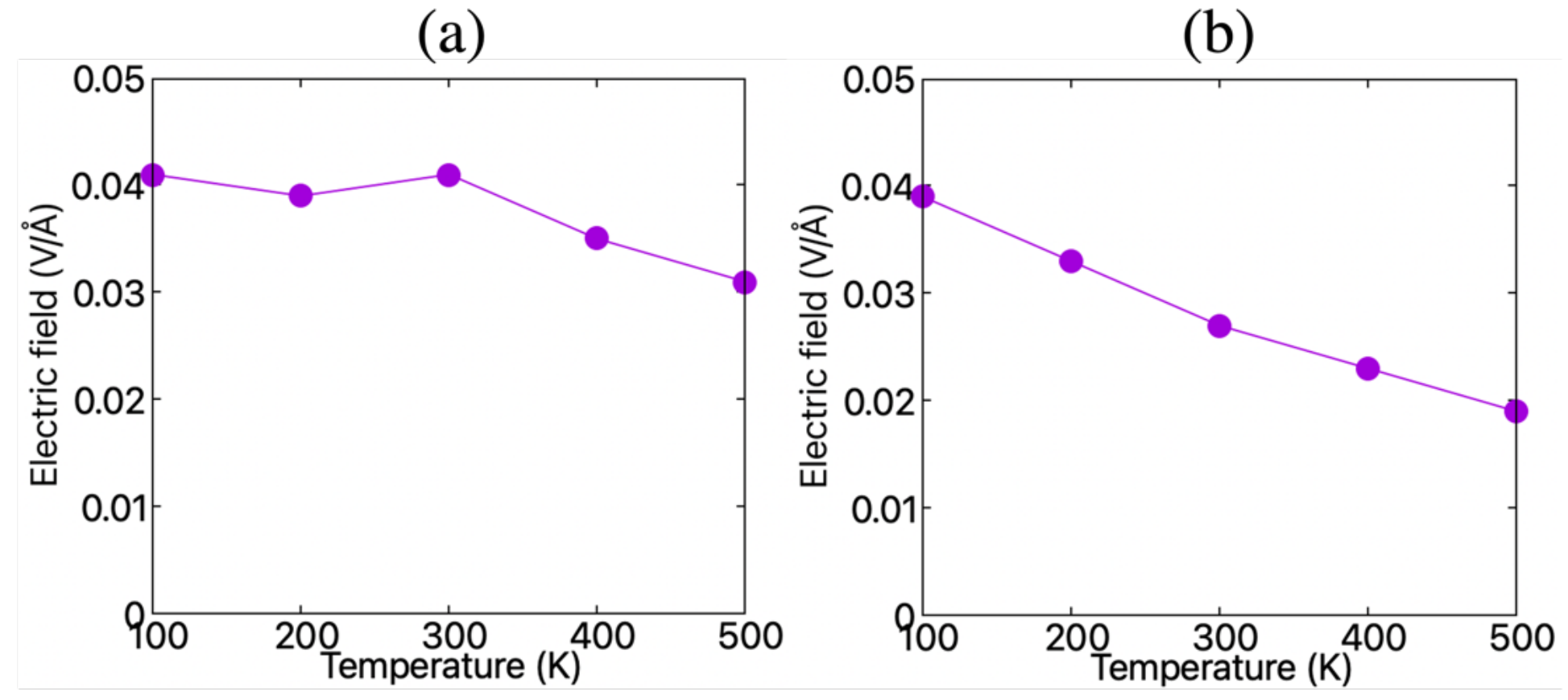

Figure 4 | Coercive field as a function of temperature for $Ba_3I_6$ (a) and $Cs_2PdC_2$ (b), obtained from MD simulations based on the MLP.

## Methods

### Density functional calculations.

First-principles calculations were performed within the framework of density functional theory using the Vienna Ab initio Simulation Package[64] (VASP). The projector augmented-wave[65,66] (PAW) approach was adopted to describe the electron–ion interactions, and the exchange–correlation effects were treated within the generalized gradient approximation in the Perdew–Burke–Ernzerhof[67] (PBE) form. The kinetic energy cutoff for the plane-wave expansion was chosen to be 520 eV, ensuring well-converged total energies and forces. The band gap for $Ba_3I_6$ and $Cs_2PdC_2$ are calculated from HSE functional[63]. The Brillouin zone integration was carried out using a Monkhorst–Pack k-point sampling scheme, with mesh densities of 8 × 4 × 4, 7 × 7 × 7, 9 × 9 × 5, and 12 × 6 × 4 for the primitive cells of $Ba_3I_6$, $Cs_2PdC_2$, $Ag_2I_2$, and $Na_4Pd_2S_4$, respectively. Atomic coordinates and lattice parameters were fully optimized until the residual forces on each atom were below 0.01 eV/Å, while the total energy convergence threshold was set to $10^{-6}$ eV.

**High-throughput screening.**

Materials were selected from the Materials Project database[56] with the following criteria: experimentally reported crystal structures, band gaps exceeding 0.1 eV, fewer than 50 atoms per unit cell, and no more than four distinct atomic species. There are 26,103 structures that satisfy these conditions. For these initial structures (L1), we generate the corresponding final structures (L2) using the implementation in PASP[68], and determine the transition paths between them using Crystmatch[53,54]. Along these paths, the energy barriers are calculated using the solid-state NEB method[55], as implemented in TSASE, based on the universal machine-learning potential model UMA[57]. After this procedure, we identify 68 QFEs and 100 conventional ferroelectrics, all with energy barriers below 400 meV per formula unit.

**MLP with Born effective charges (BECs)**

To simulate the response of materials to external electric fields, we trained the DREAM model [69], which simultaneously predicts the potential energy and the BECs of each atom based on the atomic configuration. This enables the calculation of atomic forces under an electric field. All first-principles data used for training were obtained using the VASP package. The structures in the training set were generated via ab initio molecular dynamics (AIMD), based on the L1 and L2 phases as well as the transition paths identified by the nudged elastic band (NEB) method. The BECs for each configuration were calculated using density functional perturbation theory (DFPT).

The datasets for the $Ba_3I_6$ and $Cs_2PdC_2$ systems consist of 5,087 and 2,220 structures, respectively. These data were split into training, validation, and test sets with a ratio of approximately 80:10:10. The mean absolute errors (MAEs) and root mean squared errors (RMSEs) for various labels on the test set are summarized in Table S5 of SM. Parity plots of the energy and BECs for the test set are shown in Fig. S5 of SM. Furthermore, we evaluated the model's performance in predicting the energy and BECs along the phase transition path from L1 to L2, as shown in Fig. S6 and Fig. S7 of SM, respectively. These validation results demonstrate that our MLP accurately predicts the properties of these systems under external electric fields.

## MD simulations

Molecular dynamics simulations were performed using the LAMMPS software, where the atomic forces under an electric field were provided by the trained DREAM model. The simulations for the $Ba_3I_6$ and $Cs_2PdC_2$ systems were conducted using $5 \times 5 \times 10$ (2,250 atoms) and $6 \times 6 \times 12$ (2,160 atoms) supercells, respectively. Periodic boundary conditions were applied in all three directions. The integration time step was set to 1 fs to ensure numerical stability and energy conservation, and the isothermal–isobaric (NPT) ensemble was employed.

To determine the critical electric field for each system, we incrementally increased the external electric field at various temperatures, ensuring that the system reached equilibrium at each field strength. The evolution of the system's dipole moment during this process was recorded, as shown in Fig. S8 of SM. The critical electric field at a given temperature was identified as the field strength at which the dipole moment began to increase significantly.

## Data Availability

The main data supporting the findings of this work are available within this paper. Extra data are available upon reasonable request from the corresponding authors.

## Code Availability

The computer code used for numerical calculation and data processing are available from the corresponding authors upon reasonable request.

## Acknowledgements

S. D. and H. X. are supported by the National Key R&D Program of China (Grant No. 2022YFA1402901), NSFC (Grants No. 12188101), and the Guangdong Major Project of the Basic and Applied Basic Research (Future functional materials under extreme conditions–Project No. 2021B0301030005), Shanghai Science and Technology Program (No. 23JC1400900), Shanghai Pilot Program for Basic Research—FuDan University 21TQ1400100 (23TQ017), the robotic AI-Scientist platform of Chinese Academy of Science, and New Cornerstone Science Foundation. W.L. and L.B. acknowledge the support from the Vannevar Bush Faculty Fellowship (VBFF) Grant No. N00014-20-1–2834 from the Department of Defense, the ARO Grant No. W911NF-21–1-0113, and the MonArk NSF Quantum Foundry supported by the National Science Foundation Q-AMASE-i Program under NSF Award No. DMR-1906383.

**Author Contributions**

W. L. and S. D. performed the calculations. H. X. and L. B. conceived and supervised the project. All authors contributed to the analysis and writing of the manuscript.

**Competing Interests**

The authors declare no competing interests.